\documentclass{svmult}
\usepackage{epsfig,graphicx} % figures
\usepackage[bottom]{footmisc}% places footnotes at page bottom
\usepackage{natbib,url}      %RR additions
\bibpunct{(}{)}{;}{a}{}{,}   %RR reference punctuation
\def\cite#1{\citealp{#1}}    %RR restore old astroncite \cite command
\def\authorindex#1{}  %RA to be redefined by editor at insertion into book

\begin{document}

%%\setcounter{page}{1}
%RA to insert and reset to actual page number for your Astro-PH upload

%RR file: rr-assp-defs.tex = extra ASSP definitions by Rob Rutten
%RR last: Dec 25 2008 
%RR note: ?? problem    %RR Rob-to-Rob   

%RR ## to be adapted when the volume number is known
\def\thisvolume{these proceedings}

%RR journal abbreviations
%%%%%%%%%%%%%%%%%%%%%%%%%
\def\aj{{AJ}}			
\def\araa{{ARA\&A}}		
\def\apj{{ApJ}}			
\def\apjl{{ApJ}}		
\def\apjs{{ApJS}}		
\def\ao{{Appl.\ Optics}} 
\def\apss{{Ap\&SS}}		
\def\aap{{A\&A}}		
\def\aapr{{A\&A~Rev.}}		
\def\aaps{{A\&AS}}		
\def\an{{Astron.\ Nachrichten}}
\def\aspcs{{ASP Conf.\ Ser.}}
\def\azh{{AZh}}			
\def\baas{{BAAS}}		
\def\jrasc{{JRASC}}		
\def\memras{{MmRAS}}		
\def\mnras{{MNRAS}}
\def\nat{{Nat}}		
\def\pra{{Phys.\ Rev.\ A}} 
\def\prb{{Phys.\ Rev.\ B}}		
\def\prc{{Phys.\ Rev.\ C}}		
\def\prd{{Phys.\ Rev.\ D}}		
\def\prl{{Phys.\ Rev.\ Lett}}	
\def\pasp{{PASP}}
\def\pasj{{PASJ}}		
\def\qjras{{QJRAS}}
\def\science{{Sci}}		
\def\skytel{{S\&T}}		
\def\solphys{{Solar\ Phys.}} 
\def\sovast{{Soviet\ Ast.}}  
\def\ssr{{Space\ Sci.\ Rev.}}
\def\svassp{{Astrophys.\ Space Science Proc.}}
\def\zap{{ZAp}}			
\let\astap=\aap
\let\apjlett=\apjl
\let\apjsupp=\apjs

%RR astronomy and math commands copied from ASP
%%%%%%%%%%%%%%%%%%%%%%%%%%%%%%%%%%%%%%%%%%%%%%%
\def\ion#1#2{{\rm #1}\,{\uppercase{#2}}}  %RR ~>\, \sc > uc 
\def\deg{\hbox{$^\circ$}}
\def\sun{\hbox{$\odot$}}
\def\earth{\hbox{$\oplus$}}
\def\la{\mathrel{\hbox{\rlap{\hbox{\lower4pt\hbox{$\sim$}}}\hbox{$<$}}}}
\def\ga{\mathrel{\hbox{\rlap{\hbox{\lower4pt\hbox{$\sim$}}}\hbox{$>$}}}}
\def\sq{\hbox{\rlap{$\sqcap$}$\sqcup$}}
\def\arcmin{\hbox{$^\prime$}}
\def\arcsec{\hbox{$^{\prime\prime}$}}
\def\fd{\hbox{$.\!\!^{\rm d}$}}
\def\fh{\hbox{$.\!\!^{\rm h}$}}
\def\fm{\hbox{$.\!\!^{\rm m}$}}
\def\fs{\hbox{$.\!\!^{\rm s}$}}
\def\fdg{\hbox{$.\!\!^\circ$}}
\def\farcm{\hbox{$.\mkern-4mu^\prime$}}
\def\farcs{\hbox{$.\!\!^{\prime\prime}$}}
\def\fp{\hbox{$.\!\!^{\scriptscriptstyle\rm p}$}}
\def\micron{\hbox{$\mu$m}}
\def\onehalf{\hbox{$\,^1\!/_2$}}	
\def\onethird{\hbox{$\,^1\!/_3$}}
\def\twothirds{\hbox{$\,^2\!/_3$}}
\def\onequarter{\hbox{$\,^1\!/_4$}}
\def\threequarters{\hbox{$\,^3\!/_4$}}
\def\ubv{\hbox{$U\!BV$}}		
\def\ubvr{\hbox{$U\!BV\!R$}}		
\def\ubvri{\hbox{$U\!BV\!RI$}}		
\def\ubvrij{\hbox{$U\!BV\!RI\!J$}}		
\def\ubvrijh{\hbox{$U\!BV\!RI\!J\!H$}}		
\def\ubvrijhk{\hbox{$U\!BV\!RI\!J\!H\!K$}}		
\def\ub{\hbox{$U\!-\!B$}}		
\def\bv{\hbox{$B\!-\!V$}}		
\def\vr{\hbox{$V\!-\!R$}}		
\def\ur{\hbox{$U\!-\!R$}}

%%%%%%%%%%%%%%%%%%%%%%%%%%%%%%%%%%%%%%%%%%%%%%%%%%%%%%%%%%%%%%%%%%%%%%%%%%%%
%RR RJR additional commands
%%%%%%%%%%%%%%%%%%%%%%%%%%%%%%%%%%%%%%%%%%%%%%%%%%%%%%%%%%%%%%%%%%%%%%%%%%%%

%RR -- non-bullet item marker in itemize list 
\def\labelitemi{{\bf --}}  

%RR -- latin abbreviations
\def\rmit#1{{\it #1}}              %% italics (RR style, Kluwer)
\def\rmit#1{{\rm #1}}              %% redefine for ASP, A&A, ApJ, Springer??
\def\etal{\rmit{et al.}}           %% use \etal\ for space behind it        
\def\etc{\rmit{etc.}}           
\def\ie{\rmit{i.e.,}}              %% , required (Webster 1681)
\def\eg{\rmit{e.g.,}}              %% , required (Webster 1681)
\def\cf{cf.}                       %% no Latin, always Roman (Webster 1686)
\def\viz{\rmit{viz.}}
\def\vs{\rmit{vs.}}

%RR -- mathematical
\def\rot{\hbox{\rm rot}}
\def\div{\hbox{\rm div}}
\def\lesssim{\mathrel{\hbox{\rlap{\hbox{\lower4pt\hbox{$\sim$}}}\hbox{$<$}}}}
\def\gtrsim{\mathrel{\hbox{\rlap{\hbox{\lower4pt\hbox{$\sim$}}}\hbox{$>$}}}}
\def\dif{\: {\rm d}}                       %% differential d with space
\def\ep{\:{\rm e}^}                        %% e^ with space and roman e
\def\dash{\hbox{$\,-\,$}}                  %% math-like hyphen
\def\is{\!=\!}                             %% = in text for tighter spacing

%RR --stellar stuff
\def\starname#1#2{${#1}$\,{\rm {#2}}}  %% \starname{\alpha}{Cen~A} 
\def\Teff{\hbox{$T_{\rm eff}$}}   

%RR -- units (in addition to the ASP ones above)
\def\kms{\hbox{km$\;$s$^{-1}$}}
\def\Mxcm{\hbox{Mx\,cm$^{-2}$}}    %% no 2, damn tex

%RR -- magnetic field 
\def\Bapp{\hbox{$B_{\rm app}$}}    %% apparent flux density, Lites convention

%RR -- oscillations
\def\komega{($k, \omega$)}                 %% k - omega 
\def\kf{($k_h,f$)}                         %% f - k_h
\def\VminI{\hbox{$V\!\!-\!\!I$}}           %% V-I
\def\IminI{\hbox{$I\!\!-\!\!I$}}           %% I-I
\def\VminV{\hbox{$V\!\!-\!\!V$}}           %% V-V
\def\Xt{\hbox{$X\!\!-\!t$}}                %% X-t

%RR -- atomic levels
%%      use:    \level 3s3p 3Pe
%%              \level 3s$^2$ {1,3}P{e,o}
%%              \level {} 3Ge
\def\level #1 #2#3#4{$#1 \: ^{#2} \mbox{#3} ^{#4}$}   

%RR -- some spectral species
\def\specchar#1{\uppercase{#1}}    %% to be redefined for A&A = \sc
\def\AlI{\mbox{Al\,\specchar{i}}}  %% use \AlI\ for space behind it
\def\BI{\mbox{B\,\specchar{i}}} 
\def\BII{\mbox{B\,\specchar{ii}}}  
\def\BaI{\mbox{Ba\,\specchar{i}}}  
\def\BaII{\mbox{Ba\,\specchar{ii}}} 
\def\CI{\mbox{C\,\specchar{i}}} 
\def\CII{\mbox{C\,\specchar{ii}}} 
\def\CIII{\mbox{C\,\specchar{iii}}} 
\def\CIV{\mbox{C\,\specchar{iv}}} 
\def\CaI{\mbox{Ca\,\specchar{i}}} 
\def\CaII{\mbox{Ca\,\specchar{ii}}} 
\def\CaIII{\mbox{Ca\,\specchar{iii}}} 
\def\CoI{\mbox{Co\,\specchar{i}}} 
\def\CrI{\mbox{Cr\,\specchar{i}}} 
\def\CriI{\mbox{Cr\,\specchar{ii}}} 
\def\CsI{\mbox{Cs\,\specchar{i}}} 
\def\CsII{\mbox{Cs\,\specchar{ii}}} 
\def\CuI{\mbox{Cu\,\specchar{i}}} 
\def\FeI{\mbox{Fe\,\specchar{i}}} 
\def\FeII{\mbox{Fe\,\specchar{ii}}} 
\def\FeIX{\mbox{Fe\,\specchar{ix}}}
\def\FeX{\mbox{Fe\,\specchar{x}}}
\def\FeXVI{\mbox{Fe\,\specchar{xvi}}}
\def\FrI{\mbox{Fr\,\specchar{i}}}
\def\HI{\mbox{H\,\specchar{i}}} 
\def\HII{\mbox{H\,\specchar{ii}}} 
\def\Hmin{\hbox{\rmH$^{^{_{\scriptstyle -}}}$}}      %% H^min, elegant
\def\Hemin{\hbox{{\rm He}$^{^{_{\scriptstyle -}}}$}} %% He^min, idem
\def\HeI{\mbox{He\,\specchar{i}}} 
\def\HeII{\mbox{He\,\specchar{ii}}} 
\def\HeIII{\mbox{He\,\specchar{iii}}} 
\def\KI{\mbox{K\,\specchar{i}}} 
\def\KII{\mbox{K\,\specchar{ii}}} 
\def\KIII{\mbox{K\,\specchar{iii}}} 
\def\LiI{\mbox{Li\,\specchar{i}}} 
\def\LiII{\mbox{Li\,\specchar{ii}}} 
\def\LiIII{\mbox{Li\,\specchar{iii}}} 
\def\MgI{\mbox{Mg\,\specchar{i}}} 
\def\MgII{\mbox{Mg\,\specchar{ii}}} 
\def\MgIII{\mbox{Mg\,\specchar{iii}}} 
\def\MnI{\mbox{Mn\,\specchar{i}}} 
\def\NI{\mbox{N\,\specchar{i}}}
\def\NaI{\mbox{Na\,\specchar{i}}}
\def\NaII{\mbox{Na\,\specchar{ii}}}
\def\NaIII{\mbox{Na\,\specchar{iii}}} 
\def\NiI{\mbox{Ni\,\specchar{i}}} 
\def\NiII{\mbox{Ni\,\specchar{ii}}}
\def\NiIII{\mbox{Ni\,\specchar{iii}}} 
\def\OI{\mbox{O\,\specchar{i}}} 
\def\OVI{\mbox{O\,\specchar{vi}}}
\def\RbI{\mbox{Rb\,\specchar{i}}} 
\def\SII{\mbox{S\,\specchar{ii}}} 
\def\SiI{\mbox{Si\,\specchar{i}}} 
\def\SiII{\mbox{Si\,\specchar{ii}}} 
\def\SrI{\mbox{Sr\,\specchar{i}}}
\def\SrII{\mbox{Sr\,\specchar{ii}}}
\def\TiI{\mbox{Ti\,\specchar{i}}} 
\def\TiII{\mbox{Ti\,\specchar{ii}}} 
\def\TiIII{\mbox{Ti\,\specchar{iii}}} 
\def\TiIV{\mbox{Ti\,\specchar{iv}}} 
\def\VI{\mbox{V\,\specchar{i}}} 
\def\HtwoO{\mbox{H$_2$O}}        %% H2O %RR TeX doesn't accept numbers alas
\def\Otwo{\mbox{O$_2$}}          %% O2

%RR -- hydrogen spectrum features
\def\Halpha{\mbox{H\hspace{0.1ex}$\alpha$}} %% \Halpha\ for space behind it
\def\Ha{\mbox{H\hspace{0.2ex}$\alpha$}}
\def\Hbeta{\mbox{H\hspace{0.2ex}$\beta$}}
\def\Hgamma{\mbox{H\hspace{0.2ex}$\gamma$}}
\def\Hdelta{\mbox{H\hspace{0.2ex}$\delta$}}
\def\Hepsilon{\mbox{H\hspace{0.2ex}$\epsilon$}}
\def\Hzeta{\mbox{H\hspace{0.2ex}$\zeta$}}
\def\Lyalpha{\mbox{Ly$\hspace{0.2ex}\alpha$}}
\def\Lybeta{\mbox{Ly$\hspace{0.2ex}\beta$}}
\def\Lygamma{\mbox{Ly$\hspace{0.2ex}\gamma$}}
\def\Lycont{\mbox{Ly\hspace{0.2ex}{\small cont}}}
\def\Baalpha{\mbox{Ba$\hspace{0.2ex}\alpha$}}
\def\Babeta{\mbox{Ba$\hspace{0.2ex}\beta$}}
\def\Bacont{\mbox{Ba\hspace{0.2ex}{\small cont}}}
\def\Paalpha{\mbox{Pa$\hspace{0.2ex}\alpha$}}
\def\Bralpha{\mbox{Br$\hspace{0.2ex}\alpha$}}

%RR -- Na D
\def\NaD{\mbox{Na\,\specchar{i}\,D}}    %% use \NaD\ for space behind it
\def\NaDone{\mbox{Na\,\specchar{i}\,\,D$_1$}}
\def\NaDtwo{\mbox{Na\,\specchar{i}\,\,D$_2$}}
\def\NaID{\mbox{Na\,\specchar{i}\,\,D}}
\def\NaIDone{\mbox{Na\,\specchar{i}\,\,D$_1$}}
\def\NaIDtwo{\mbox{Na\,\specchar{i}\,\,D$_2$}}
\def\Done{\mbox{D$_1$}}
\def\Dtwo{\mbox{D$_2$}}

%RR -- Mg b 
\def\Mgbone{\mbox{Mg\,\specchar{i}\,b$_1$}}
\def\Mgbtwo{\mbox{Mg\,\specchar{i}\,b$_2$}}
\def\Mgbthree{\mbox{Mg\,\specchar{i}\,b$_3$}}
\def\MgIb{\mbox{Mg\,\specchar{i}\,b}}
\def\MgIbone{\mbox{Mg\,\specchar{i}\,b$_1$}}
\def\MgIbtwo{\mbox{Mg\,\specchar{i}\,b$_2$}}
\def\MgIbthree{\mbox{Mg\,\specchar{i}\,b$_3$}}

%RR -- Ca II H & K 
\def\CaIIK{\mbox{Ca\,\specchar{ii}\,K}}       %% use \CaIIK\ for space
\def\CaIIH{\mbox{Ca\,\specchar{ii}\,H}}
\def\CaIIHK{\mbox{Ca\,\specchar{ii}\,H\,\&\,K}}
\def\HK{\mbox{H\,\&\,K}}
\def\Kthree{\mbox{K$_3$}}      %% numbers not permitted, alas
\def\Hthree{\mbox{H$_3$}}
\def\Ktwo{\mbox{K$_2$}}
\def\Htwo{\mbox{H$_2$}}
\def\Kone{\mbox{K$_1$}}     
\def\Hone{\mbox{H$_1$}}     
\def\KtwoV{\mbox{K$_{2V}$}}
\def\KtwoR{\mbox{K$_{2R}$}}
\def\KoneV{\mbox{K$_{1V}$}}
\def\KoneR{\mbox{K$_{1R}$}}
\def\HtwoV{\mbox{H$_{2V}$}}
\def\HtwoR{\mbox{H$_{2R}$}}
\def\HoneV{\mbox{H$_{1V}$}}
\def\HoneR{\mbox{H$_{1R}$}}

%RR -- Mg II h & k 
\def\hk{\mbox{h\,\&\,k}}
\def\kthree{\mbox{k$_3$}}    
\def\hthree{\mbox{h$_3$}}
\def\ktwo{\mbox{k$_2$}}
\def\htwo{\mbox{h$_2$}}
\def\kone{\mbox{k$_1$}}     
\def\hone{\mbox{h$_1$}}     
\def\ktwoV{\mbox{k$_{2V}$}}
\def\ktwoR{\mbox{k$_{2R}$}}
\def\koneV{\mbox{k$_{1V}$}}
\def\koneR{\mbox{k$_{1R}$}}
\def\htwoV{\mbox{h$_{2V}$}}
\def\htwoR{\mbox{h$_{2R}$}}
\def\honeV{\mbox{h$_{1V}$}}
\def\honeR{\mbox{h$_{1R}$}}
  
%RA use these as needed but don't change nor add!
%RA   if you really need private /def's then define them here
%RA   so that we can inspect them and put into rr-assp-defs
%RA   if really useful - and not upsetting any other paper

\title*{Deep-Focus Diagnostics of Sunspot Structure}
%RA capitalize the nouns, try to fit on one line

%%\titlerunning{YourStuffHere}
%RA capitalize the nouns
%RA only if needed for a too long title, not in this case

\author{H.~Moradi\inst{1, 3}
        \and
        S.~M.~Hanasoge\inst{2, 3}
        }
\authorindex{Moradi,~H.}
\authorindex{Hanasoge,~S.~M.}

\institute{School of Mathematical Sciences, Monash University, Australia.
           \and
           Max-Planck-Institut f\"ur Sonnensystemforschung, Katlenburg-Lindau, Germany.
           \and
           Indian Institute for Astrophysics, Bangalore, India.
           }
\maketitle

\setcounter{footnote}{0}  %RR Springer forgot this one (and much more)

%%%%%%%%%%%%%%%%%%%%%%%%%%%%%%%%%%%%%%%%%%%%%%%%%%%%%%%%%%%%%%%%%%%%%%%%%%%%
\begin{abstract} 
Following on from the recent results of \citet{mhc2008}, we employ two established numerical forward models (a 3D ideal MHD solver and MHD ray theory), in conjunction with time-distance helioseismology to probe the lateral extent of wave-speed perturbations produced in regions of strong, near-surface magnetic fields. We aim to continue our comparative studies of the forward models by complimenting our previous surface-focused travel-time measurements through the application of a common midpoint deep-focusing scheme that avoids the use of oscillation signals within the sunspot region. The idea here is to also test MHD ray theory for possible application to future inverse methods. 
\end{abstract}
%%%%%%%%%%%%%%%%%%%%%%%%%%%%%%%%%%%%%%%%%%%%%%%%%%%%%%%%%%%%%%%%%%%%%%%%%%%%
%RA no keywords please

%%%%%%%%%%%%%%%%%%%%%%%%%%%%%%%%%%%%%%%%%%%%%%%%%%%%%%%%%%%%%%%%%%%%%%%%%%%%
\section{Introduction}      \label{yourname-sec:introduction}
%%%%%%%%%%%%%%%%%%%%%%%%%%%%%%%%%%%%%%%%%%%%%%%%%%%%%%%%%%%%%%%%%%%%%%%%%%%%
%RA do not capitalize nouns, only the first word
In \citet{mhc2008} we utilized two recently developed numerical MHD forward models, in conjunction with surface-focused (i.e., centre-to-annulus) time-distance measurements, to produce numerical models of travel-time inhomogeneities in a simulated sunspot atmosphere. The resulting artificial travel-time perturbation profiles clearly demonstrated the overwhelming influence that MHD physics, as well as phase-speed and frequency filtering, have on local helioseismic measurements in the vicinity of sunspots. 

However, there are numerous caveats associated with surface-focused time-distance measurements that use oscillation signals within the sunspot region, as the use of such oscillation signals is now known to be the primary source of most surface effects in sunspot seismology.  These surface effects can be essentially categorized into two groups. The first revolves around the degree to which observations made within the sunspot region are contaminated by magnetic effects (e.g., \citealp{braun1997,lb2005,schunkeretal2005,bb2006,cr2007,mhc2008}), while the second concerns the degree to which atmospheric temperature stratification in and around regions may affect the absorption line used to make measurements of the Doppler velocity (e.g., \citealp{rajaguruetal2006,rajaguruetal2007}). 

There have been attempts in the past to circumvent such problems by adopting a time-distance measurement geometry known as ``deep-focusing''  which avoids the use of data from the central area of the sunspot by only cross-correlating the oscillation signal of waves that have a first-skip distance larger than the diameter of the sunspot (e.g., \citealp{duvall1995,braun1997,zk2006,rajaguru2008}). In this analysis, we follow up on the comparative study presented in \citet{mhc2008} by using our two established forward models, in conjunction with a deep-focusing scheme known as the ``common midpoint'' (CMP) method to probe the sub-surface dynamics of our artificial sunspot.  

\section{The Flux Tube and Forward Models}
\begin{figure}
\centering
%\begin{tabular}{cc}
%\begin{center}
%\epsscale{1.0}
\includegraphics[width=28pc]{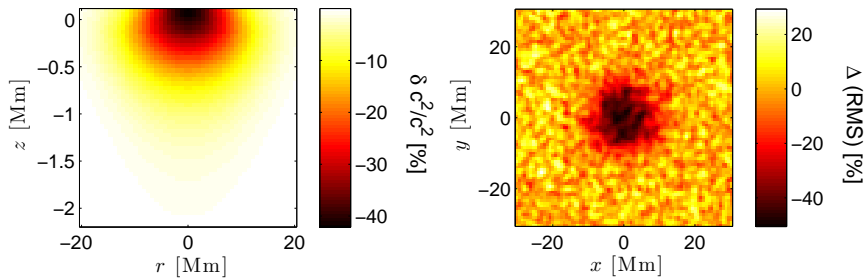}
%\includegraphics[width=16pc]{cs_ws.eps}
%\end{center}
%\end{tabular}
\caption{Some properties of the model sunspot atmosphere. Left panel: the near-surface thermal/sound-speed perturbation profile, shown as a function of sound-speed squared. Right panel:  shows a Doppler power map, normalized to the quiet Sun.}
\label{fig:figs1} 
\end{figure}

The background stratification of our model atmosphere is given by an adiabatically stable, truncated polytrope \citep{bogdanetal1996}, smoothly connected to an isothermal atmosphere. The truncated polytrope is described by: index $m=2.15$, reference pressure $p_0=1.21\times10^5$ g cm$^{-1}$ s$^{-2}$ and reference density $\rho_0=2.78\times10^{-7}$ g cm$^{-3}$, The flux tube (peak field strength of 3~kG) is modeled by an axisymmetric magnetic field geometry based on the \citet{st1958} self-similar solution. The derived MHS sunspot model achieves a consistent sound-speed decrease (see Figure \ref{fig:figs1}), with a peak reduction of $\sim 45\%$ at the surface ($z=0$) and less than $1\%$ at  $z=-2$~Mm, while the one-layered wave-speed enhancement is also confined to the near-surface layers, approaching $\sim 200\%$ at the surface and around less than $0.5\%$ at $z=-2$~Mm. 

The two forward models presented in \citet{mhc2008} are again used for our analysis. The first forward model integrates the linearized ideal MHD wave equations according to the recipe of \citet{hanasoge2008}, where waves are excited via a pre-computed deterministic source function that acts on the vertical momentum equation. To simulate the suppression of granulation related wave sources in a sunspot (e.g., \citealp{hanasogeetal2007}), the source activity is muted in a circular region of 10 Mm radius. The simulations produce artificial line-of-sight (Doppler) velocity data cubes, extracted at a height of 200~km above the photosphere, in effect, mimicking Michelson Doppler Imager (MDI) Dopplergrams. The data cubes have dimensions of $200\times200$~Mm$^2$ $\times$ $512$ minutes, with a cadence of 1 minute and a spatial resolution of $0.718$~Mm. Figure~\ref{fig:figs1} depicts a normalized power map derived from the simulated Doppler velocity measurements. 

The second forward model employs the MHD ray tracer of \citet{mc2008}, where (2D) ray propagation is modeled through solving the governing equations of the ray paths derived using the zeroth order eikonal approximation and the magneto-acoustic dispersion relation. It should be noted that neither forward model accounts for the presence of sub-surface flows.

\section{Common Midpoint Deep-Focusing}
\begin{figure}
\centering
%\begin{tabular}{cc}
%\begin{center}
%\epsscale{1.0}
\includegraphics[width=28pc]{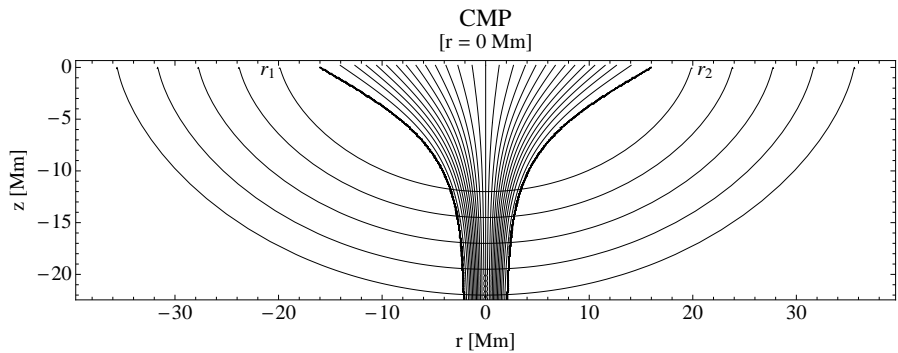}
%\includegraphics[width=16pc]{cs_ws.eps}
%\end{center}
%\end{tabular}
\caption{An illustration of the CMP deep-focus geometry indicating the range of rays used for this study. The CMP method measures the travel time at the point on the surface located at the half-way point between a source ($\bf{r_1}$) and receiver ($\bf{r_2}$). For the above rays, the CMP is located on the central axis of the spot ($r=0$~Mm). }
\label{fig:cmp} 
\end{figure}

Often utilized in geophysics applications such as multichannel seismic acquisition \citep{shearer1999}, the CMP method measures the travel time at the point on the surface halfway between the source and receiver (see Figure \ref{fig:cmp}). Cross-correlating numerous source-receiver pairs in this manner results in the method being mostly sensitive to a small region in the deep interior surrounding the lower turning point of the ray. A re-working of this method has been applied to helioseismic observations by \citet{duvall2003}, and has the obvious advantage of allowing one to study the wave-speed structure directly beneath sunspots without using the oscillation signals inside the perturbed region. 

Our method for measuring time-distance deep-focus travel times is somewhat similar to the approach undertaken by \citet{braun1997} and  \citet{duvall2003}. First, the annulus-to-annulus cross-covariances (e.g., between oscillation signals located between two points on the solar surface, a source at $\bf{r_1}$ and a receiver at $\bf{r_2}$, as illustrated in Figure \ref{fig:cmp}) are derived by dividing each annulus ($\bf{\Delta}=|\bf{r_2}-\bf{r_1}|$), into two semi-annuli (each being one data pixel wide) and cross-correlating the average signals in these two semi-annuli. Then, to further increase the signal-to-noise ratio (SNR), we average the cross-covariances over 3 distances, respectively slightly smaller than, and larger than, $\bf{\Delta}$. In the end, the 5 (mean) distances chosen ($\bf{\Delta}=~$42.95, 49.15, 55.35, 61.65 and 68~Mm respectively) are large enough to ensure that we only sample waves with a first-skip distance greater than the diameter of the sunspot at the surface ($\sim40$~Mm). 

Due to the oscillation signal at any location being a superposition of a large number of waves of different travel distances, the cross-covariances are very noisy and need to be phase-speed filtered first in the Fourier domain, using a Gaussian filter for each travel distance. The application of appropriate phase-speed filters isolates waves that travel desired skip distances, meaning that even though we average over semi-annuli, the primary contribution to the cross covariances is from these waves. In addition to the phase-speed filters, we also apply an $f$-mode filter that removes the $f$-mode ridge completely (as it is of no interest to us in this analysis), and we also apply Gaussian frequency filters centred at $\omega=$~3.5, 4.0 and 5.0~mHz with $\delta\omega=$~0.5~mHz band-widths, to study frequency dependencies of travel times (e.g., \citealp{bb2008,mhc2008}). To extract the required travel times, the cross-covariances are fitted by two Gabor wavelets \citep{kd1997}: one for the positive times, one for the negative times. 

\begin{figure}
\centering
%\epsscale{1.0}
\includegraphics[width=28pc]{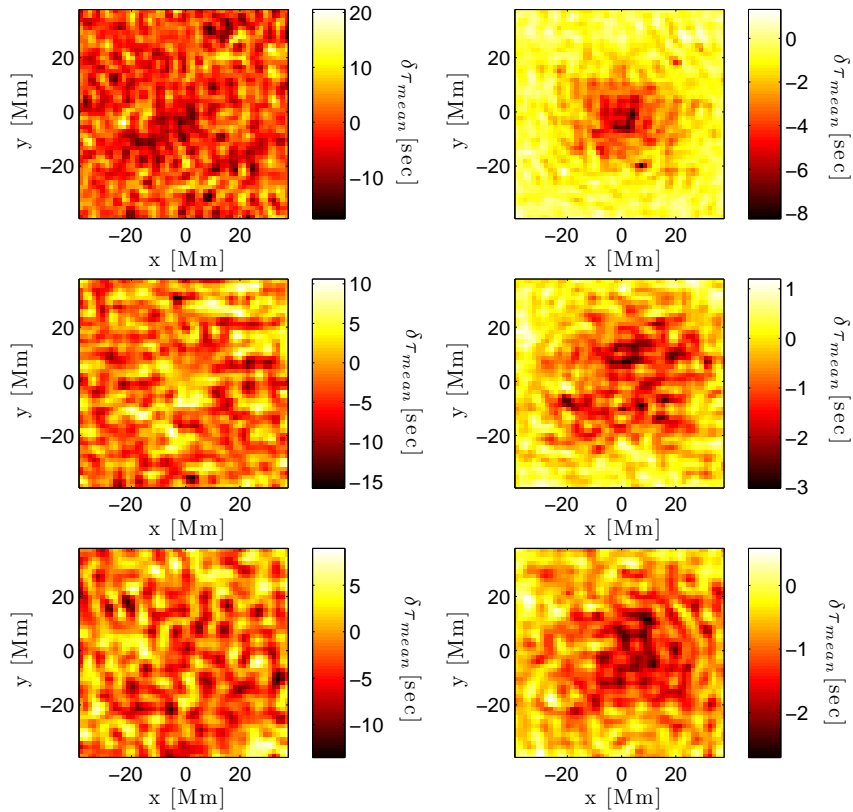}
\caption{Examples of (phase-speed filtered) CMP mean travel-time perturbation ($\delta\tau_{mean}$) maps for $\bf{\Delta}=42.95$ (top), $\bf{\Delta}=49.15$ (middle) and $\bf{\Delta}=61.65$~Mm (bottom). Left panels: before realization noise subtraction. Right panels: after subtraction. A frequency filter centred at 5.0~mHz has been applied to the data.}
\label{fig:tdmaps} 
\end{figure}

Even after significant filtering and averaging, the extracted CMP travel times are still inundated with noise. This is certainly an ever-present complication in local helioseismology as there is a common expectation (with all local helioseismic methods and inversions) of worsening noise and resolution with depth. Realization noise associated with stochastic excitation of acoustic waves can significantly impair our ability to analyse the true nature of travel-time shifts on the surface (and by extension, also affect our interpretation of sub-surface structure). But, as we have full control over the wave excitation mechanism and source function, we have the luxury of being able to apply realization noise subtraction to improve the SNR and obtain statistically significant travel-time shifts from the deep-focus measurements. This is accomplished in the same manner as \citet{hdc2007}, i.e., by performing two separate simulations, one with the perturbation (i.e., the sunspot simulation), and another without (i.e., the quiet simulation). We then subtract the travel times of the quiet data from its perturbed counterpart (see Figure~\ref{fig:tdmaps}), allowing us to achieve an excellent SNR.  

Finally, in order to compare theory with simulations, we also estimate deep-focusing time shifts using the MHD ray tracer of \citet{mc2008}. The single-skip magneto-acoustic rays are propagated from the inner (lower) turning point of their trajectories at a prescribed frequency (see e.g., Figure \ref{fig:cmp}). These rays do not undergo any additional filtering as the required range of horizontal skip distances are simply obtained by altering the depth at which the rays are initiated from. The resulting mean (phase) travel-time shifts ($\delta\tau_{mean}$) derived from both forward models are presented in Figures \ref{fig:tdmaps} and \ref{fig:figs3}.

\section{Results and Discussion}  
\begin{figure}
\centering
%\begin{tabular}{cc}
%\begin{center}
%\epsscale{1.0}
\includegraphics[width=28pc]{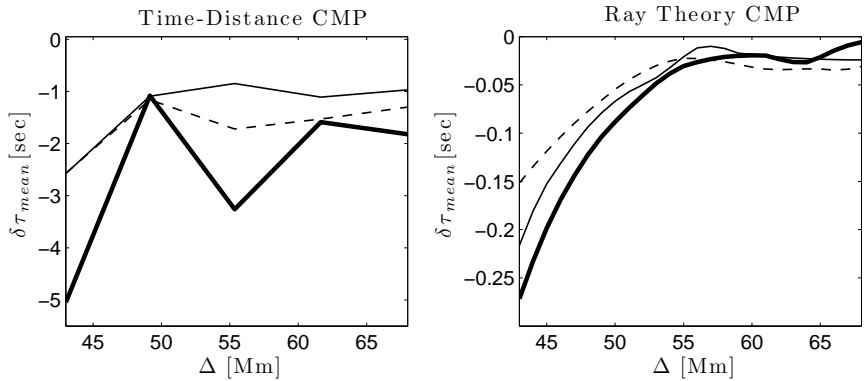}
%\includegraphics[width=16pc]{cs_ws.eps}
%\end{center}
%\end{tabular}
\caption{Observed CMP travel-time shifts as a function of wave/ray travel distance ($\bf{\Delta}$). Left panel: umbral averages of the CMP time shifts derived from time-distance analysis of the simulated data. Right panel: ray theory CMP travel-time shifts derived from rays propagated at various depths and with a CMP at $r=0$~Mm. Light sold lines are indicative of frequency filtering centred at 3.5~mHz, dashed lines indicate 4.0~mHz and bold solid lines indicate 5.0~mHz.}
\label{fig:figs3} 
\end{figure}

A number of travel time maps derived from the time-distance analysis, both before and after noise subtraction, are presented in Figure~\ref{fig:tdmaps}.  The impact of realization noise subtraction is self-evident in these figures as it is only after removing the background noise that we are able to detect statistically significant travel-time shifts. The umbral averages of these time shifts are shown in Figure~\ref{fig:figs3}. The $\delta\tau_{mean}$ range from a couple of seconds at 3.5 and 4.0~mHz, to around five seconds at 5.0~mHz. However, even though the size of the measured time shifts are significant, there is no clear frequency dependence associated with the them. As we are only using waves outside of the perturbed region, surface effects can be effectively ruled out as the cause of the time shifts. 

It is worth noting that linear inversions of surface-focused travel time maps of actual observations have suggested a two-layered wave-speed structure below sunspots - a wave-speed decrease of $\sim10-15\%$ down to a depth of $\sim3-4$~Mm, followed by a wave-speed enhancement, reportedly detected down to depths of $\sim17-25$~Mm below the surface \citep{kds2000,cbk2006}. However, with our forward model clearly prescribing both a shallow wave- and sound-speed perturbation profile (Figure \ref{fig:figs1}), it is hard to fathom that the time-distance $\delta\tau_{mean}$ we are observing can be associated with some kind of anomalous deep sub-surface perturbation. In fact, both the sound-speed decrease and wave-speed enhancement at such depths registers at less than one-tenth of one percent and the value of the plasma $\beta$ is in the range of $\sim7-18\times10^3$ -- in all likelihood not significant enough to produce a 3-5 second travel-time perturbation. In order to try and identify the root cause of these apparent travel-time shifts, it is useful to compare the time-distance CMP measurements with those derived from MHD ray theory in Figure \ref{fig:figs3}. 

The ray theory CMP $\delta\tau_{mean}$ clearly appear to be significantly smaller at all frequencies, with all observed time shifts registering at less than half a second. Certainly, these time shifts are more in line with our expectations given the absence of any significant deep sound/wave-speed perturbation. But, we must bear in mind the differences between the two forward models before drawing our conclusions. With regards to helioseismic travel times, \citet{bogdan1997} has emphasized that they are not only sensitive to the local velocity field along the ray path, but also to conditions in the surrounding medium -- a clear consequence of wave effects. As such, wave-like behaviour needs to be considered when interpreting travel times -- something which ray theory does not clearly account for, resulting in possible underestimation of deep-focus travel times. 

On the other hand, we must also consider the effects of phase-speed filtering (which is absent in the ray theory calculations) on the time-distance measurements. If we look closely at the time-distance $\delta\tau_{mean}$ maps in Figure~\ref{fig:tdmaps}, we notice that they are somewhat smeared in appearance, with the central sunspot region becoming increasingly sprawled-out across the map as we increase $\bf{\Delta}$. This behaviour is most likely a consequence of both the phase-speed filtering (i.e, the size of the central frequency filter, the filter width, etc., see \citealp{cb2006}), and the averaging scheme applied to the cross-correlations -- both of which are a necessity in order to improve the SNR in time-distance calculations. These effects, combined with the delocalized nature of the CMP measurements, may also introduce spurious travel-time shifts. However, further testing and control simulations are required to confirm this. 

\section{Conclusion}
At the present time, it is sufficient to say that we do not have a definitive diagnosis with regards to the above-discussed differences in the size of the deep-focus time shifts produced by the two forward models. It may well be that we are applying ray theory to regimes where it may be seriously limited. On the other hand, the very same could be said about local time-distance analysis! Whatever the case may be, these preliminary results have certainly provided us with the motivation to conduct further time-distance studies using the CMP method. 

The direct (and indirect) effects of phase-speed filtering on deep-focus measurements, derived from both simulations and real data, also warrants a more detailed examination, as any artifact produced by the filtering process is likely to be even more pronounced for phase-speed filtered MDI data, where we do not yet have the luxury of realization noise subtraction. These issues are something that we hope to address in the very near future with some ongoing comparative studies.

%RA figure always between blank lines = paragraph 
%============================================================================
%\begin{figure}  
%RA no location specifier [t] or such please
%  \centering
%  \includegraphics[width=\textwidth]{\figspath/yourname-figXXX}
%RA or smaller by declaring cm; the text width = 11.8cm 
%  \caption[]{\label{yourname-fig:XXX}
%RA don't forget the [] 
%
%  YourStuffHere
% 
%}\end{figure}
%===========================================================================
%%%%%%%%%%%%%%%%%%%%%%%%%%%%%%%%%%%%%%%%%%%%%%%%%%%%%%%%%%%%%%%%%%%%%%%%%%%%
\begin{acknowledgement}
The authors express their gratitude to the Indian Institute for Astrophysics for their support and warm hospitality during their stay in Bangalore which made this work possible. The authors also thank Charlie Lindsey and Paul Cally for insightful discussions regarding various aspects of this work. HM also acknowledges the travel support provided by the Astronomical Society of Australia. 
\end{acknowledgement}

%%RA figure always between blank lines = paragraph 
%%============================================================================
%\begin{figure}  
%%RA no location specifier [t] or such please
%  \centering
%  \includegraphics[width=\textwidth]{\figspath/yourname-figXXX}
%%RA or smaller by declaring cm; the text width = 11.8cm 
%  \caption[]{\label{yourname-fig:XXX}
%%RA don't forget the [] 
%%
%  YourStuffHere
%% 
%}\end{figure}
%%===========================================================================

%
%%%%%%%%%%%%%%%%%%%%%%%%%%%%%%%%%%%%%%%%%%%%%%%%%%%%%%%%%%%%%%%%%%%%%%%%%%%%%
%\section{Conclusion}                   \label{yourname-sec:conclusion}
%%%%%%%%%%%%%%%%%%%%%%%%%%%%%%%%%%%%%%%%%%%%%%%%%%%%%%%%%%%%%%%%%%%%%%%%%%%%%
%%RA do not capitalize nouns, only the first word

%YourStuffHere

%
%%%%%%%%%%%%%%%%%%%%%%%%%%%%%%%%%%%%%%%%%%%%%%%%%%%%%%%%%%%%%%%%%%%%%%%%%%%%
%\begin{acknowledgement}
%  We thank the conference organisers for a very good meeting and the
%  editors for excellent instructions. %RA -:)
%\end{acknowledgement}

%%%%%%%%%%%%%%%%%%%%%%%%%%%%%%%%%%%%%%%%%%%%%%%%%%%%%%%%%%%%%%%%%%%%%%%%%%%%
%% References
%%%%%%%%%%%%%%%%%%%%%%%%%%%%%%%%%%%%%%%%%%%%%%%%%%%%%%%%%%%%%%%%%%%%%%%%%%%%
\begin{small}

%RA Use commands as the following two to generate the bibliography
%RA automatically with BibTeX as file xxx.bbl.  

\bibliographystyle{rr-assp}       %RR hacked from aa.bst
%\bibliography{/Users/hamedmoradi/PhD/Thesis_II/Biblio}

\begin{thebibliography}{25}
\expandafter\ifx\csname natexlab\endcsname\relax\def\natexlab#1{#1}\fi

\bibitem[{{Bogdan}(1997)}]{bogdan1997}
{Bogdan}, T.~J. 1997, \apj, 477, 475

\bibitem[{{Bogdan} {et~al.}(1996){Bogdan}, {Hindman}, {Cally}, \&
  {Charbonneau}}]{bogdanetal1996}
{Bogdan}, T.~J., {Hindman}, B.~W., {Cally}, P.~S., {Charbonneau}, P. 1996,
  \apj, 465, 406

\bibitem[{{Braun}(1997)}]{braun1997}
{Braun}, D.~C. 1997, \apj, 487, 447

\bibitem[{{Braun} \& {Birch}(2006)}]{bb2006}
{Braun}, D.~C. {Birch}, A.~C. 2006, \apjl, 647, L187

\bibitem[{{Braun} \& {Birch}(2008)}]{bb2008}
{Braun}, D.~C. {Birch}, A.~C. 2008, \solphys, 251, 267

\bibitem[{{Couvidat} \& {Birch}(2006)}]{cb2006}
{Couvidat}, S. {Birch}, A.~C. 2006, \solphys, 237, 229

\bibitem[{{Couvidat} {et~al.}(2006){Couvidat}, {Birch}, \&
  {Kosovichev}}]{cbk2006}
{Couvidat}, S., {Birch}, A.~C., {Kosovichev}, A.~G. 2006, \apj, 640, 516

\bibitem[{{Couvidat} \& {Rajaguru}(2007)}]{cr2007}
{Couvidat}, S. {Rajaguru}, S.~P. 2007, \apj, 661, 558

\bibitem[{{Duvall}(1995)}]{duvall1995}
{Duvall}, Jr., T.~L. 1995, in GONG 1994. Helio- and Astro-Seismology from the
  Earth and Space, eds. R.~K. {Ulrich}, E.~J. {Rhodes}, Jr., \& W.~{Dappen},
  Astronomical Society of the Pacific Conference Series, 76, 465

\bibitem[{{Duvall}(2003)}]{duvall2003}
{Duvall}, Jr., T.~L. 2003, in GONG+ 2002. Local and Global Helioseismology: the
  Present and Future, ed. H.~{Sawaya-Lacoste}, ESA Special Publication, 517,
  259

\bibitem[{{Hanasoge}(2008)}]{hanasoge2008}
{Hanasoge}, S.~M. 2008, \apj, 680, 1457

\bibitem[{{Hanasoge} {et~al.}(2008){Hanasoge}, {Couvidat}, {Rajaguru}, \&
  {Birch}}]{hanasogeetal2007}
{Hanasoge}, S.~M., {Couvidat}, S., {Rajaguru}, S.~P., {Birch}, A.~C. 2008,
  \mnras, 1291

\bibitem[{{Hanasoge} {et~al.}(2007){Hanasoge}, {Duvall}, \&
  {Couvidat}}]{hdc2007}
{Hanasoge}, S.~M., {Duvall}, Jr., T.~L., {Couvidat}, S. 2007, \apj, 664, 1234

\bibitem[{{Kosovichev} \& {Duvall}(1997)}]{kd1997}
{Kosovichev}, A.~G. {Duvall}, T.~L. 1997, in ASSL Vol. 225: SCORe'96 : Solar
  Convection and Oscillations and their Relationship, 241

\bibitem[{{Kosovichev} {et~al.}(2000){Kosovichev}, {Duvall}, \&
  {Scherrer}}]{kds2000}
{Kosovichev}, A.~G., {Duvall}, T.~L.~.~J., {Scherrer}, P.~H. 2000, \solphys,
  192, 159

\bibitem[{{Lindsey} \& {Braun}(2005)}]{lb2005}
{Lindsey}, C. {Braun}, D.~C. 2005, \apj, 620, 1107

\bibitem[{{Moradi} \& {Cally}(2008)}]{mc2008}
{Moradi}, H. {Cally}, P.~S. 2008, \solphys, 251, 309

\bibitem[{{Moradi} {et~al.}(2009){Moradi}, {Hanasoge}, \& {Cally}}]{mhc2008}
{Moradi}, H., {Hanasoge}, S.~M., {Cally}, P.~S. 2009, \apjl, 690, L72

\bibitem[{{Rajaguru}(2008)}]{rajaguru2008}
{Rajaguru}, S.~P. 2008, ArXiv e-prints, 802

\bibitem[{{Rajaguru} {et~al.}(2006){Rajaguru}, {Birch}, {Duvall}, {Thompson},
  \& {Zhao}}]{rajaguruetal2006}
{Rajaguru}, S.~P., {Birch}, A.~C., {Duvall}, Jr., T.~L., {Thompson}, M.~J.,
  {Zhao}, J. 2006, \apj, 646, 543

\bibitem[{{Rajaguru} {et~al.}(2007){Rajaguru}, {Sankarasubramanian}, {Wachter},
  \& {Scherrer}}]{rajaguruetal2007}
{Rajaguru}, S.~P., {Sankarasubramanian}, K., {Wachter}, R., {Scherrer}, P.~H.
  2007, \apjl, 654, L175

\bibitem[{{Schl{\"u}ter} \& {Temesv{\'a}ry}(1958)}]{st1958}
{Schl{\"u}ter}, A. {Temesv{\'a}ry}, S. 1958, in Electromagnetic Phenomena in
  Cosmical Physics, ed. B.~{Lehnert}, IAU Symposium, 6,  263

\bibitem[{{Schunker} {et~al.}(2005){Schunker}, {Braun}, {Cally}, \&
  {Lindsey}}]{schunkeretal2005}
{Schunker}, H., {Braun}, D.~C., {Cally}, P.~S., {Lindsey}, C. 2005, \apjl, 621,
  L149

\bibitem[{{Shearer}(1999)}]{shearer1999}
{Shearer}, P. 1999, {Introduction to Seismology} (Introduction to Seismology,
  by Peter Shearer, pp.~272.~ISBN 0521660238.~Cambridge, UK: Cambridge
  University Press, September 1999.)

\bibitem[{{Zhao} \& {Kosovichev}(2006)}]{zk2006}
{Zhao}, J. {Kosovichev}, A.~G. 2006, \apj, 643, 1317

\end{thebibliography}

%RA Ignore the error messages that BibTeX may give.

%RA After manuscript completion, insert the final .bbl file
%RA below and comment the next two statements out.

%RA We advise you to add unique labels (such as yourname-) to
%RA all bibtex labels, both in the \cite commands above and after the ]{
%RA symbol combinations below.  Otherwise identical ADS labels from another
%RA author with worse \bibitems then yours may muck up your good work.

\end{small}

\end{document}